\documentclass[twocolumn,preprintnumbers,amsmath,revsymb]{revtex4-1}

\usepackage{graphicx}
\usepackage{dcolumn}
\usepackage{bm}
\usepackage{feynmf}
\usepackage{verbatim}

\newcommand{\p}{\partial×}

\newcommand{\ds}{\displaystyle×}

\newcommand{\ket}{\rangle}
\newcommand{\bra}{\langle×}

\begin{document}

\title{Neutron-Antineutron Oscillations in a Warped Extra Dimension}

\author{Peter T. Winslow}
\email{pwinslow@phas.ubc.ca}
\affiliation{Department of Physics and Astronomy, University of British Columbia, Vancouver, BC, V6T 1Z1, Canada}
\affiliation{Theory Group, TRIUMF, 4004 Wesbrook Mall, Vancouver, BC, Canada, V6T 2A3}
\author{John N. Ng}
\email{misery@triumf.ca}
\affiliation{Theory Group, TRIUMF, 4004 Wesbrook Mall, Vancouver, British Columbia, Canada, V6T 2A3}

\begin{abstract}
\indent \indent We investigate neutron-antineutron oscillations in the Randall-Sundrum warped extra dimensional scenario. The four dimensional effective strengths of the relevant operators that induce the oscillations are calculated up to an arbitrary coupling along with their corresponding enhancements due to QCD 1-loop running effects. We find that the $\Delta B = 2$ operators can be geometrically suppressed without fine tuning to within current experimental limits with a warped down four dimensional mass scale which can be as low as a fraction of a TeV.
\end{abstract}

\maketitle

\section{Introduction}

The standard model (SM) of particle physics has, for 30 years, enjoyed unequaled success in describing the results of particle physics experiments. It is, however, not an entirely satisfactory theory due to the fact that it has, to date, left many unanswered fundamental questions. In particular, it provides no explanation for the many different hierarchies which have been built into it. The most famous of these being the electroweak-Planck hierarchy problem in which, due to the ultra-violet sensitivity of the Higgs mass, the massive separation between the Planck scale and the electroweak scale is considered to be unnatural. One particularly appealing solution to this problem is the Randall-Sundrum (RS) model~\cite{Randall:1999ee,Davoudiasl:1999tf,Flachi:2001bj,Randall:2001gb}. Within the context of this model the large hierarchy arises due to the warping of a compactified Anti-deSitter (AdS) extra dimensional geometry. This warping manifests itself as a warping factor which exponentially suppresses the mass scales within the theory, creating an effective hierarchy. Another appealing feature  of the RS model is its ability to explain the SM fermion mass hierarchies with the same mechanism which explains the electroweak-Planck hierarchy~\cite{Huber:2000ie,Gherghetta:2000qt,Kaplan:2001ga}. By promoting all SM fermions to bulk fields the fermion mass hierarchies are explained in terms of the fermion geography within the warped extra dimensional space. In such a scenario, the five dimensional (5D) fermion fields are Dirac fields whose wave function localization in the extra dimension is completely characterized by a single $\mathcal{O}(1)$ $c$ parameter. By using a $Z_2$ orbifold projection or equivalently by choosing appropriate boundary conditions on the UV and IR branes one can project out the chiral zero modes. The SM fermions are identified with these  chiral zero modes of the bulk fermions and  they have exponential wavefunction profiles in the extra dimension. The effective Yukawa couplings depend heavily on the wave function overlap of the corresponding fermion wavefunctions with the Higgs, which is situated on the TeV brane in the extra dimension. Heavy fermions are localized near the IR brane and thereby have a large overlap with the Higgs field, while light fermions are localized closer to the UV brane. In this particular class of RS flavour models the SM gauge symmetry is promoted to a bulk symmetry.  The 4D Yukawa couplings can then all be taken to be $\mathcal{O}(1)$ while the $c$ parameters can all be determined by fitting the fermion masses and their mixing parameters~\cite{Chang:2008zx}. 

One can then ask about the nature of higher mass dimension operators within the context of the RS model, such as those corresponding to proton decay and neutron-antineutron ($n$-$\overline{n}$) oscillation~\cite{Gopalakrishna:2002es,Mohapatra:2009wp,Dubbers:1991bh,Huber:2000ie}. If no extra symmetry forbids these operators they will be suppressed by some mass scale which is close to or exceeds the UV completion scale of the RS model. If one simply takes this to be the Planck scale then this would be sufficient to satisfy the experimental constraints however the exact same warping mechanism which reduces the Planck scale to the electroweak scale acts to reduce this mass scale suppression as well. 

It is well known that proton decay is a problem for the RS model~\cite{Huber:2000ie}. In order to properly suppress the relevant operators for proton decay it is necessary to maintain large separations between the quarks and leptons in the extra dimension however successful mass configurations for these fields do not allow for such large separations. As opposed to accepting unnaturally small dimensionless couplings for these operators it is thought that there exists an extra symmetry which will forbid these operators entirely however the exact nature of this symmetry is as yet unknown. The simplest solution is to introduce a $U_X(1)$ symmetry where $X$ could denote the total baryon number (B), lepton number (L), or their difference (B-L) which is currently understood to be only an accidental symmetry of the SM gauge group. Discrete symmetries of the $Z_N$ type have also been suggested.

It is thought that $n$-$\overline{n}$ oscillations could present yet another problem for models of this type due to the fact that the corresponding operators contain only quarks of similar mass scales and therefore similar localizations within the extra dimension~\cite{Mohapatra:2009wp}. In the current work we analyze the effective strength of the six quark operators which induce $n$-$\overline{n}$ oscillations in the warped RS model assuming that there is no symmetry which a priori forbids them. For example, the introduction of a $U_L(1)$ or $Z_3$ symmetry would have no effect on the operators which induce $n$-$\overline{n}$ oscillations but does forbid the operators which induce proton decay. Discrete symmetries have also been used to study Dirac neutrinos in warped models \cite{Chang:2009mv}. Previous investigations of the effective strength of $n$-$\overline{n}$ oscillation operators within the context of the 6D \mbox{Arkani-Hamed--Dimopoulos--Dvali} (ADD) have yielded a lower bound on the mass scale suppression in the observable range $M_X \gtrsim (45 - 100)$ TeV~\cite{Nussinov:2001rb}. \\

The paper is organized as follows: Section~\ref{sec:formalism} briefly reviews the treatment of fermions on the 5D AdS$_5$ background and the problems with proton decay in the RS model. Section~\ref{sec:oscillation} introduces the relevant operators which induce the $n$-$\overline{n}$ oscillations and presents the corresponding 4D effective Wilson coefficients. In section~\ref{sec:QCD} we present the calculated enhancement of the strength of these coefficients due to SM QCD 1-loop renormalization group (RG) running effects which are expected to be larger than other gauge interactions and in section~\ref{sec:conclusion} we discuss our results.  

\section{Formalism}
\label{sec:formalism}

\indent \indent This section serves to define our conventions and notation. The 5D space is mapped by coordinates $x^A = (x^\mu, \phi)$ where the fifth dimension is compactified with size $r_c$ and $\phi \in [-\pi, \pi]$. In order to embed a 4D Minkowski spacetime within a slice of 5D anti-deSitter space (AdS$_5$) with curvature $k$ the points $(x^\mu, \phi)$ and $(x^\mu, -\phi)$ are identified. This creates an $S^1/Z_2$ orbifold with fixed points $(x^\mu, 0)$ and $(x^\mu, \pi)$. The metric components of the warped non-factorizable geometry are given by the line element~\cite{Randall:1999ee}

\begin{equation}
ds^2 = G_{A B} dx^A dx^B = e^{-2 \sigma( \phi ) } \eta_{\mu \nu} dx^\mu dx^\nu - r_c^2 d \phi^2
\end{equation}

where $x^\mu$ are the coordinates on the four dimensional hypersurfaces of constant $\phi$ with the Minkowski metric $\eta_{\mu \nu}$ and $\sigma(\phi) = k r_c |\phi|$. Two three branes, called the ultraviolet (UV) and the infrared (IR) branes, are placed at the orbifold fixed points $\phi=0$ and $\phi=\pi$ respectively. The parameters $k$ and $1/r_c$ are assumed to be on the order of the Planck scale while the product $k r_c$ is chosen to be $\sim 12$ to solve the hierarchy problem. 

Working in the low energy effective field theory we can write down the free field action for a massive fermion $\Psi(x,\phi)$ in the RS background as 

\begin{equation}
S = \int d^4 x d \phi \sqrt{|G|} \left[ \frac{i}{2×} E^A_a \overline{\Psi} \gamma^a \overleftrightarrow{\partial_A} \Psi - m \; \textrm{sgn} (\phi) \overline{\Psi} \Psi \right]
\end{equation}
 
where $E^A_a = \textrm{diag} ( e^\sigma, e^\sigma, e^\sigma, e^\sigma, 1/r_c )$ is the inverse f$\ddot{\textrm{u}}$nfbein, $\gamma^a = (\gamma^\mu, i \gamma^5)$ and $m \; \textrm{sgn} (\phi)$ is the mass. The mass must be dependent on the position within the extra dimension in order for the mass term to remain invariant under the action of the $Z_2$ orbifold. Due to the diagonal nature of the metric the spin connection term does not contribute to the action~\cite{Grossman:1999ra}. Any gauge interactions are included by simply replacing $\partial_A$ with the relevant covariant derivative. 

The normalized KK mode expansion of the chiral fermion fields is chosen to be 

\begin{equation}
\psi_{L,R} (x,\phi) = \frac{e^{3/2 \sigma}}{\sqrt{r_c}×} \sum_n \psi_{n_{L,R}} (x) \chi_{n_{L,R}} (\phi)
\end{equation}

where the eigenfunctions are orthonormal such that 

\begin{align}
\int_{-\pi}^\pi d \phi \; \chi_{n_L}^* \chi_{m_L} = \int_{-\pi}^\pi d \phi \; \chi_{n_R}^* \chi_{m_R} = \delta_{n m}
\end{align}

All SM fields are associated with the zero modes of the expansion for which the normalized eigenfunctions, determined from solving the zero mode field equations, are given by 

\begin{equation}
\chi_{0_{L,R}} = \sqrt{\frac{k r_c (1/2 \pm c_{L,R})}{e^{2 k r_c \pi (1/2 \pm c_{L,R})} - 1×}} \; e^{(1/2 \pm c_{L,R}) \sigma}
\end{equation}

where the normalization factor insures a canonically normalized kinetic term in the four dimensional effective theory. It is exactly the $c_{L,R}=m/k$ parameters which control the localization of the fermion wavefunctions in the extra dimension, i.e; $c_R<1/2$ ($c_R>1/2$) corresponds to a closer proximity to the UV (IR) brane for the right handed zero mode while $c_L<-1/2$ ($c_L>-1/2$) corresponds to a closer proximity to the IR (UV) brane for the left handed zero mode. With the Higgs localized at the IR brane the effective SU$_L$(2) $\times$ U$_Y$(1) invariant fermion-Higgs Yukawa interaction with order one couplings, $y_{i j}$, in 5D is given by 

\begin{equation}
\int d^4 x d \phi \sqrt{G} \delta ( \phi - \pi) \left( \frac{y_{i j}}{k r_c}\overline{\Psi}_{i L} (x, \phi) \Phi (x) \Psi_{j R} (x, \phi) + h.c. \right)
\end{equation}

where $\Psi_L$ and $\Psi_R$ are the fermion $SU_L(2)$ doublet and singlet respectively and $\Phi$ is the Higgs $SU_L(2)$ doublet. Integrating out the extra dimensional dependence yields the effective mass matrix 

\begin{equation}
M_{i j} = y_{i j} \frac{v_w}{\sqrt{2}} \mathcal{N}_{i_L} \mathcal{N}_{j_R} e^{k r_c \pi \left(1 + c_{i_L} - c_{j_R} \right)} 
\end{equation}

where $v_w = v e^{-k r_c \pi} = 247$ GeV and we have defined the normalization factor

\begin{equation}
\mathcal{N}_{i_{L,R}} = \sqrt{\frac{\frac{1}{2×} \pm c_{i_{L,R}} }{e^{2 k r_c \pi \big( \frac{1}{2×} \pm c_{i_{L,R}} \big) } - 1}}
\end{equation}

for notational brevity. The nature of this effective mass matrix is such that the heavy fermions must be interpreted as being localized near the IR brane and therefore have a large overlap with the Higgs field while the light fermions must be localized near the UV brane yielding a small overlap with the Higgs. Another consequence is that only $\mathcal{O}(1)$ differences in the $c$ parameters are then needed to generate the observed large effective fermion mass hierarchies without any fine tuning of the dimensionless Yukawa couplings. In particular, using only $\mathcal{O}(1)$ differences between the nine $c$ parameters in the quark sector, it is possible to reproduce the entire set of quark masses and mixing angles. To this extent, there have been a number of parameter sets put forward which fit the observed data \cite{Chang:2008zx,Moreau:2006np,Ledroit:2007ik}. 

Since the RS model is itself considered to be an effective theory one can introduce higher dimensional operators in the same vein as in other extensions of the SM. Some of the more well known examples of these types of operators are those responsible for proton decay such as

\begin{equation}
\int d^4 x d \phi \sqrt{G} \frac{1}{M^3×} \left( g_1 QQQL + g_2 U^c U^c D^c E^c \right)
\end{equation}

where the fields $Q$, $L$, $U$, $D$, and $E$ are the bulk versions of the corresponding SM fields. Since these operators are understood as arising from physics above the RS UV cutoff one can conservatively take the mass scale suppression $M$ to be on the order of the Planck mass. Integrating out the 5D degrees of freedom reveals the effective strength with which the zero modes of the above fields will induce proton decay. Although the effective strength of these operators receives a suppression from the resulting exponential overlaps between the various fields the Planck scale suppression is warped down and replaced by $M e^{-k r \pi}$. The end result of these two competing effects is that there is not enough suppression from the resulting wavefunction overlap to prevent proton decay within the current limits and we must therefore concede either the fine tuning of the dimensionless couplings or the introduction of a convenient symmetry (such as total lepton number or the above mentioned $Z_3$) which will forbid the operator entirely~\cite{Huber:2000ie,Gopalakrishna:2002es}. Neither of the above mentioned symmetries forbid the operators which induce $n$-$\overline{n}$ oscillations and the question of whether or not the RS model can inherently provide the needed suppression for these transitions from geometry alone is the subject of the present work.

\section{${\bf n}$-$\overline{{\bf n}}$ Oscillations}
\label{sec:oscillation}

The time evolution of an initially slow moving beam of neutrons is described by the following Schroedinger equation involving the simple 2 $\times$ 2 Hamiltonian

\begin{equation}
i \hbar \frac{\p}{\p t×} 
\left(
\begin{array}{c}
n \\
\overline{n}
\end{array}
\right) = 
\left(
\begin{array}{cc}
E_n & \delta m \\
\delta m & E_{\overline{n}}
\end{array}
\right)
\left(
\begin{array}{c}
n \\
\overline{n}
\end{array}
\right) 
\end{equation}

where $\delta m = \langle \overline{n} | \mathcal{H}_{eff} | n \rangle$ parameterizes the underlying physics describing the oscillation~\cite{Mohapatra:2009wp,Dubbers:1991bh,Kuo:1980ew}. The probability of finding an antineutron after some time $t$ is then given by

\begin{equation}
\label{nnbaramp}
|\langle \overline{n} | n(t) \rangle|^2 = \frac{4 \delta m^2}{\Delta E^2 + 4 \delta m^2×} \sin^2 \left( \sqrt{\Delta E^2 + 4 \delta m^2} t \right)
\end{equation}

where $\Delta E = E_n - E_{\overline{n}}$ is the energy splitting. Experimental limits from reactors and matter instability have produced limits on the off-diagonal components of the Hamiltonian of $\delta m \leq \;  $.75$\times$10$^{-32} \; $GeV$^{-1}$ and $\delta m \leq \; $.6$\times$10$^{-32} \; $GeV$^{-1}$ respectively~\cite{Baldoceolin:1994,Takita:1986zm,Berger:1989gw}.

The effective Hamiltonian is given by $\mathcal{H}_{eff} = \sum_i g_i \mathcal{O}_i (x,\phi)$ where each effective operator, $\mathcal{O}_i (x,\phi)$, is a SU$_c$(3) $\times$ SU$_L$(2) $\times$ U$_Y$(1) gauge invariant six-quark operator. Any general $\Delta B = 2$ operator which contributes to $n$-$\overline{n}$ oscillations and is constructed from non-scalar Lorentz invariant quark couplings can be converted to an equivalent operator constructed strictly from scalar Lorentz invariant quark couplings via Fierz transformations~\cite{Basecq:1983hi,Rao:1982gt}. There are four linearly independent operators of this type which are given by~\cite{Nussinov:2001rb}

\begin{equation}
\mathcal{O}_1 = \big( u_R^{\alpha T} C u_R^\beta \big) \big( d_R^{\gamma T} C d_R^\delta \big) \big( d_R^{\lambda T} C d_R^\tau \big) T^s_{\alpha \beta \gamma \delta \lambda \tau}
\end{equation}

\begin{align}
\mathcal{O}_2 = \big( u_R^{\alpha T} C d_R^\beta \big) \big( u_R^{\gamma T} C d_R^\delta \big) \big( d_R^{\lambda T} C d_R^\tau \big) T^s_{\alpha \beta \gamma \delta \lambda \tau}
\end{align}

\begin{align}
\mathcal{O}_3 = \big( Q_L^{i \alpha T} C Q_L^{j \beta} \big) \big( u_R^{\gamma T} C d_R^\delta \big) \big( d_R^{\lambda T} C d_R^\tau \big) \epsilon_{i j} T^a_{\alpha \beta \gamma \delta \lambda \tau}
\end{align}

\begin{align}
& \mathcal{O}_4 = \big( Q_L^{i \alpha T} C Q_L^{j \beta} \big) \big( Q_L^{k \gamma T} C Q_L^{l \delta} \big) \big( d_R^{\lambda T} C d_R^\tau \big) \epsilon_{i j} \epsilon_{k l} T^a_{\alpha \beta \gamma \delta \lambda \tau} \notag \\
\end{align}

The round brackets are meant to imply the contraction of spinor indices, $C$ is the charge conjugation operator, and Greek and Latin indices represent SU$_c$(3) and SU$_L$(2) degrees of freedom respectively. The color tensors contract the $SU_c$(3) indices into color singlet combinations in two different ways given by~\cite{Rao:1983sd}  

\begin{equation}
T^s_{\alpha \beta \gamma \delta \lambda \tau} = \epsilon_{\tau \beta \delta} \; \epsilon_{\lambda \alpha \gamma } + \epsilon_{\tau \alpha \gamma} \; \epsilon_{\lambda \beta \delta} + \epsilon_{\tau \alpha \delta} \; \epsilon_{\lambda \beta \gamma} + \epsilon_{\tau \beta \gamma} \; \epsilon_{\lambda \alpha \delta}
\end{equation}

\begin{equation}
T^a_{\alpha \beta \gamma \delta \lambda \tau} = \epsilon_{\tau \alpha \beta} \; \epsilon_{\lambda \gamma \delta} + \epsilon_{\tau \gamma \delta} \; \epsilon_{\lambda \alpha \beta}
\end{equation}

where the first tensor is symmetric about the interchanges $(\alpha, \beta)$, $(\gamma, \delta)$, $(\lambda, \tau)$, $(\alpha \beta, \gamma \delta)$, $(\alpha \beta, \lambda \tau)$, $(\gamma \delta, \lambda \tau)$ while the second is anti-symmetric about the interchanges $[\alpha, \beta]$ and $[\gamma, \delta]$ and symmetric about $(\lambda, \tau)$ and $(\alpha \beta, \gamma \delta)$. These operators can all be easily generalized to 5D by replacing the fermion fields with the corresponding bulk fields. In 5D the coefficients associated with each operator have mass dimension -7 so we can rewrite them as $g_i = \ds \frac{C_i}{M_X^7}$ where the $C_i$'s are the dimensionless Wilson coefficients and $M_X$ is the 5D mass scale at which a detailed knowledge of the underlying physics responsible for the generation of the operators becomes indispensable. In order to obtain the effective dimensionless couplings and mass scale in the 4D theory we integrate out the extra dimensional dependence as

\begin{equation}
\label{eq:Heff}
\frac{C_i}{M_{X}^7} \int_{-\pi}^\pi d \phi \sqrt{|G|} \mathcal{O}_i (x, \phi) = \frac{C_i^{eff}}{M_{4D}^5} \mathcal{O}_i (x)
\end{equation}

where $C_i^{eff}$ is the 4D effective Wilson coefficient associated with the $i^{th}$ operator and $M_{4D}=M_X e^{-k r_c \pi}$ is the warped down 4D mass scale which is independent of the localization of the particle content in the extra dimension. The physics that generates these operators is largely unknown and highly model dependent. If they arise due to a higher dimensional grand unified theory then the mass scale $M_X$ is expected to be of order $k$, the AdS$_5$ curvature scale, although if they arise due to physics beyond the RS UV cutoff then the mass scale can be of order $1/r_c$. In our effective theory approach it is appropriate to parameterize the mass scale as $M_X = \rho k$ with $\rho$ taken to be a free parameter of the theory to be determined by experiment. The warped down 4D effective mass scale is then written as $M_{4D} = \rho k e^{-k r_c \pi}$. This is a convenient choice of parametrization since studies of precision electroweak measurements and flavour changing neutral currents imply that the lowest allowed value of the warped down curvature scale is $k e^{- k r_c \pi} = 1.65$ TeV~\cite{Chang:2008vx}. \\

Writing the zero modes of the SM quark bulk fields as 

\begin{equation}
q_{L,R} (x,\phi) = \sqrt{k} \; \mathcal{N}_{L,R} \; q_{L,R}(x) \; e^{(2 \pm c_{L,R} ) \sigma}
\end{equation}

yields the following set of effective Wilson coefficients

\begin{equation}
C_1^{eff} = C_2^{eff} = \frac{C_1 \; \mathcal{N}_{u_R}^2 \; \mathcal{N}_{d_R}^4 \; e^{k r_c \pi (3 - 2 c_{u_R} - 4 c_{d_R}) } }{ \rho^2 (4 - c_{u_R} - 2 c_{d_R} ) }
\end{equation}

\begin{equation}
C_3^{eff} = \frac{2 C_3 \; \mathcal{N}_{d_R}^{3} \; \mathcal{N}_{Q_L}^2 \; \mathcal{N}_{u_R} \; e^{k r_c \pi (3 + 2 c_{Q_L} - c_{u_R} - 3 c_{d_R}) } }{ \rho^2 (8 + 2 c_{Q_L} - c_{u_R} - 3 c_{d_R} ) }
\end{equation}

\begin{equation}
C_4^{eff} = \frac{C_4 \; \mathcal{N}_{Q_L}^4 \; \mathcal{N}_{d_R}^2 \; e^{k r_c \pi (3 + 4 c_{Q_L} - 2 c_{d_R}) } }{ \rho^2 (4 + 2 c_{Q_L} - c_{d_R} ) }
\end{equation}

The equality of $C_1^{eff}$ and $C_2^{eff}$ is due to the fact that the two operators $\mathcal{O}_1$ and $\mathcal{O}_2$ share the same overall quark content and differ only in the way in which the spinor and color indices are contracted. Furthermore, as the UV complete theory is not known the dimensionless Wilson coefficients $C_i$ are also unknown. With no loss of generality we then set all the dimensionless Wilson coefficients to unity. 

As previously mentioned, there have been a number of numerical fits made to the existing observational data taking into account not only the quark masses but also the CKM mixing angles as well. In table I we reproduce a list of three different representative configurations of $c$ parameters for the first generation of the left and right handed quarks which fit the data~\cite{Chang:2008zx}

\begin{table}[ht]
\caption{Numerical Fits of the Quark Masses with CKM Mixing Angles}
\centering
\begin{tabular}{|c|c|c|c|}
\hline
 & \multicolumn{3}{c|}{Configurations} \\
\hline
$c$ parameters & I & II & III \\
 \hline
$c_{Q_L}$ & -0.634 & -0.629 & -0.627 \\
\hline
$c_{u_R}$ & 0.664 & 0.662 & 0.518 \\
\hline
$c_{d_R}$ & 0.641 & 0.58 & 0.576 \\
\hline
\end{tabular}
\end{table}

where, due to the $SU_L(2)$ gauge symmetry, $c_{u_L}$ and $c_{d_L}$ are equal and we have denoted them both simply as $c_{Q_L}$. The matrix elements of the 4D effective six quark operators $\langle \overline{n} | \mathcal{O}_i (x) | n \rangle$ have been calculated within the context of the MIT bag model in reference~\cite{Rao:1982gt}. Averaging the results of the various fits we obtain the values $\langle \overline{n} | \mathcal{O}_1 (x) | n \rangle = -5.945 \times 10^{-5}$ GeV$^6$, $\langle \overline{n} | \mathcal{O}_2 (x) | n \rangle = 1.485 \times 10^{-5}$ GeV$^6$, $\langle \overline{n} | \mathcal{O}_3 (x) | n \rangle = -2.95 \times 10^{-5}$ GeV$^6$, and $\langle \overline{n} | \mathcal{O}_4 (x) | n \rangle = 2.22 \times 10^{-5}$ GeV$^6$. 

From Eq.(\ref{eq:Heff}) the matrix element $\langle \overline{n} | \mathcal{H}_{eff} | n \rangle$  will involve the Wilson coefficients $C^{eff}_i$ which are determined at the scale of $M_{4D}$ whereas  $n$-$\overline{n}$ oscillations take place at the neutron mass scale 1 GeV. This requires us to use the relevant RG equations to run the Wilson coefficients down to the oscillation scale. The largest contribution to the running of the Wilson coefficients will come from the QCD sector. This calculation of this contribution is the subject of the next section. 

\section{QCD running of $C^{eff}_i$}
\label{sec:QCD}

It is well known that  only the $SU_c(3)$ coupling runs significantly within the range between the TeV  scale and the neutron mass scale. We therefore restrict ourselves to calculating the $SU_c(3)$ renormalization effects only. Working to first order in $\alpha_s = \ds \frac{g_s^2}{4 \pi×}$ we find a total of 15 diagrams for each operator that must be computed not including the wavefunction renormalization diagrams. The first few generic diagrams are depicted in  FIG. 1. 

\pagebreak

\begin{figure}[ht]
\begin{centering}
\includegraphics[width=3.25in, height=1.5in, angle=0]{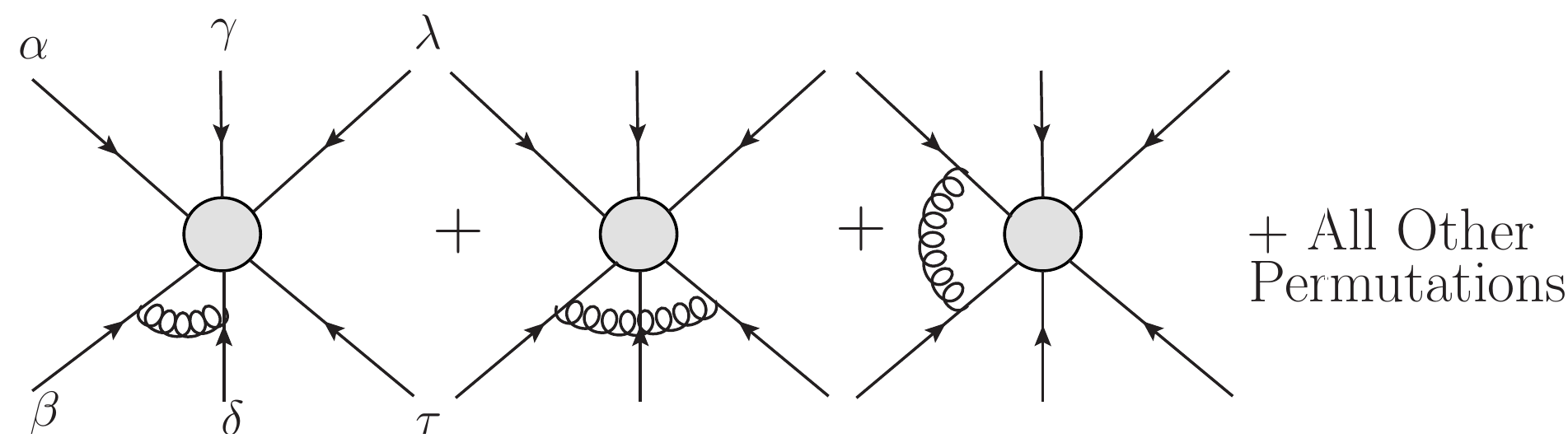}
\caption{Feynman diagrams contributing to the anomalous dimension of the effective operators $\mathcal{O}_i(x)$ due to SM gluon exchange.}
\end{centering}
\end{figure}

The renormalized effective Lagrangian is 

\begin{equation}
\mathcal{L}_{eff} = - \frac{1}{M_{4D}^5×} \sum_i \left[ C_i^{eff} \mathcal{O}_{i R} + (Z_{\mathcal{O}_i} Z_q^3 - 1) C_i^{eff} \mathcal{O}_{i R} \right]
\end{equation}

The renormalized operators $\mathcal{O}_{i R}$ are expressed in terms of the unrenormalized operators $\mathcal{O}_i$ as 

\begin{equation}
\mathcal{O}_{i R} = Z_{\mathcal{O}_i}^{-1} Z^3_q \mathcal{O}_i = Z_{\mathcal{O}_i}^{-1} \mathcal{O}_{0 i}
\end{equation}

where $\mathcal{O}_{0 i}$ are the bare operators which are independent of the renormalization scale $\mu$. Since the renormalized operator is dependent on $\mu$ via $Z_{\mathcal{O}_i}^{-1}$ the 4D effective Wilson coefficients must carry a compensating $\mu$ dependence to ensure that $\mathcal{L}_{eff}$ is independent of renormalization scale. Since the $SU_c(3)$ 1-loop running effects do not induce any operator mixing this implies that the effective Wilson coefficients each obey a simple RG equation given by

\begin{equation}
\label{RGeq}
\mu \frac{\p C_i^{eff}}{\p \mu×} + \gamma_{\mathcal{O}_i} C_i^{eff} = 0
\end{equation}
 
where $\gamma_{\mathcal{O}_i} = - Z_{\mathcal{O}_i}^{-1} \mu \ds \frac{\p Z_{\mathcal{O}_i}}{\p \mu×}$ is the anomalous dimension of the operator $\mathcal{O}_i$. Through direct calculation the counter terms for $\mathcal{O}_1$ and $\mathcal{O}_2$ are found to be equal while the same is found to be true for $\mathcal{O}_3$ and $\mathcal{O}_4$. Using the known quark wave function renormalization~\cite{Peskin}

\begin{align}
Z_q = 1 - \frac{\alpha_s}{3 \pi} \ln \left( \frac{\Lambda^2}{\mu^2} \right)
\end{align}

the independent operator renormalizations were determined to be

\begin{align}
& \; \; \; \; \; \; \; \; \; \; \; \; \; \; \; Z_{\mathcal{O}_{1,2}} = 1 \notag \\ \notag \\
& Z_{\mathcal{O}_{3,4}} = 1 + \frac{2 \alpha_s}{\pi×} \ln \left( \Lambda^2 / \mu^2 \right)
\end{align}

where $\Lambda$ is the RS UV cutoff. These operator renormalizations yield the anomalous dimensions $\gamma_{\mathcal{O}_{1,2}} = 0$ and $\gamma_{\mathcal{O}_{3,4}} = \ds \frac{4 \alpha_s (M_{4D})}{\pi×}$ where the running coupling is evaluated at the intermediate mass scale. Integrating the RG equations~(\ref{RGeq}) down to the the neutron mass scale leads to the following scaling behaviour for the effective Wilson coefficients

\begin{align}
C_{1,2}^{eff} (M_{4D}^2) = C_{1,2}^{eff} (\textrm{GeV}^2)
\end{align}

\begin{align}
C_{3,4}^{eff} (M_{4D}^2) & = C_{3,4}^{eff} (\textrm{GeV}^2) \left[ \frac{\alpha_s(\textrm{GeV}^2)}{\alpha_s(m_c^2)×} \right]^{8/9} \left[ \frac{\alpha_s(m_c^2)}{\alpha_s(m_b^2)×} \right]^{24/25} \notag \\
& \times \left[ \frac{\alpha_s(m_b^2)}{\alpha_s(m_t^2)×} \right]^{24/23} \left[ \frac{\alpha_s(m_t^2)}{\alpha_s(M_{4D}^2)×} \right]^{8/7}
\end{align}

where $m_c$, $m_b$, and $m_t$ are the masses of the charm, bottom, and top quark respectively. The full matrix element, evaluated at the neutron mass scale, which parameterizes $n$-$\overline{n}$ oscillations is then given by

\begin{widetext}
\begin{align}
\bra \overline{n} | H_{eff} | n \ket &= \frac{1}{\rho^7 (k e^{-k r_c \pi})^5×} \Bigg[ C_1^{eff} (\textrm{GeV}^2) \bigg( \bra \overline{n} | \mathcal{O}_1 | n \ket + \bra \overline{n} | \mathcal{O}_2 | n \ket \bigg) + \bigg( C_3^{eff} (\textrm{GeV}^2) \bra \overline{n} | \mathcal{O}_3 | n \ket + C_4^{eff} (\textrm{GeV}^2) \bra \overline{n} | \mathcal{O}_4 | n \ket \bigg) \notag \\
& \; \; \; \; \; \; \; \; \; \; \; \; \; \; \; \; \; \; \; \; \; \; \; \; \; \; \; \; \times \left[ \frac{\alpha_s(\textrm{GeV}^2)}{\alpha_s(m_c^2)×} \right]^{8/9} \left[ \frac{\alpha_s(m_c^2)}{\alpha_s(m_b^2)×} \right]^{24/25} \left[ \frac{\alpha_s(m_b^2)}{\alpha_s(m_t^2)×} \right]^{24/23} \left[ \frac{\alpha_s(m_t^2)}{\alpha_s(\rho k e^{-k r_c \pi})^2)×} \right]^{8/7} \Bigg]
\end{align}
\end{widetext}

where we have factored out the $1/\rho^2$ dependence from all of the effective Wilson coefficients. This leads to a simple overall $\rho$ dependence of the form

\begin{equation}
|\delta m| = |\bra \overline{n} | H_{eff} | n \ket| = \frac{1}{\rho^7×} \left( A + B \ln \rho \right)
\end{equation}

where $A$ and $B$ vary depending on which configuration is used for the $c$ parameters in the quark sector. In FIG. 2 $|\delta m|$ is plotted for each of the three configurations consistent with numerical fits to the quark masses and CKM mixing angles in TABLE I. 

\begin{figure}[ht!]
\begin{centering}
\includegraphics[width=3.5in, height=3.5in, angle=0]{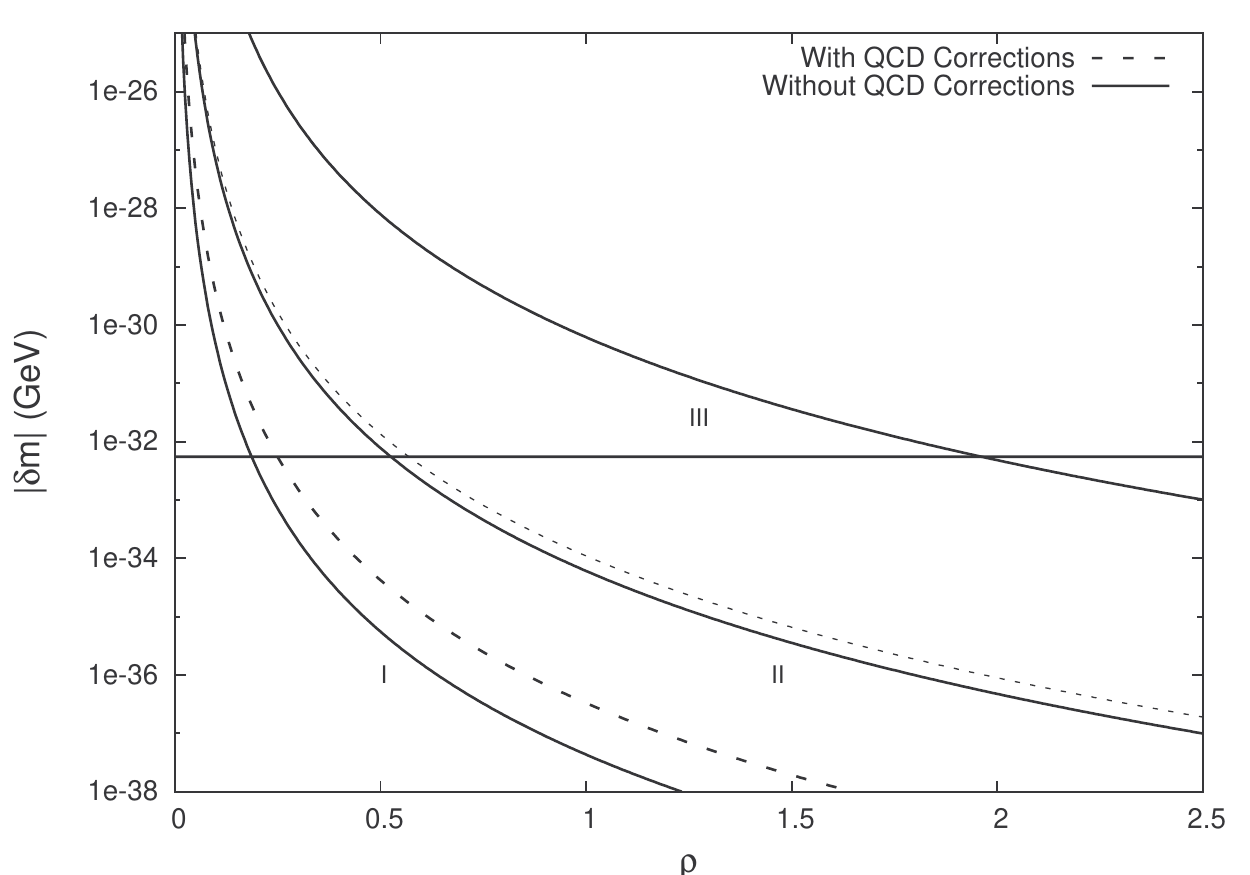}
\caption{$\delta m$ is plotted for each of the three configurations listed in TABLE I. The horizontal line represents the experimental limit $|\delta m_{exp}|$.}
\end{centering}
\end{figure}

For completeness we have included both the curves with and without the QCD 1-loop running effects. For configuration III the effective Wilson coefficients are such that $C_1^{eff}$ dominates by 2 and 5 orders of magnitude over $C_3^{eff}$ and $C_4^{eff}$ respectively. The 1-loop running effects are therefore negligible and the two curves are essentially indistinguishable on the present scale. An upper limit on $\rho$ for all three configurations can also be obtained by requiring that the value of $|\delta m|$ be less than the experimental limit $|\delta m_{exp}| = .55 \times 10^{-32}$~\cite{Nussinov:2001rb,Baldoceolin:1994,Takita:1986zm,Berger:1989gw}. Imposing this constraint for each of the three configurations leads to the following bounds: $\rho_I \gtrsim 0.240591$, $\rho_{II} \gtrsim 0.568982$, and $\rho_{III} \gtrsim 1.96185$. We can easily turn these bounds on $\rho$ into bounds on the 4D warped down effective mass scale $M_{4D}$ which are then given by $M_{4D}^I \gtrsim 0.4$ TeV, $M_{4D}^{II} \gtrsim 0.94$ TeV, and $M_{4D}^{III} \gtrsim 3.24$ TeV. This implies that only a relatively small warped down 4D mass suppression is actually needed to satisfy the currently observed experimental limits. 

\section{Conclusions}
\label{sec:conclusion}
We have investigated the effective strength of the linearly independent set of six quark operators which induce neutron-antineutron oscillations within the context of the warped Randall-Sundrum model. The overall strength of the relevant operators arose from the combination of the resultant wavefunction overlap of the six quark fields in the extra dimensional bulk, the 4D effective warped down mass scale suppression, and, to a lesser extent, QCD 1-loop running effects. 
The 4D effective warped down mass scale suppression was parameterized by a dimensionless factor in order to determine the extent of any extra suppression needed beyond the minimum allowed by flavour changing neutral current constraints. It was determined, for the quark $c$ parameter configurations listed, that the constraints on the dimensionless factor are such that the effective warped down mass scale never has to be greater than $\mathcal{O}(1)$ TeV and, in two of the three configurations, is only required to be a fraction of a TeV even with enhancements due to 1-loop running effects. The enhancements due to the QCD running were included but were determined to not have an overtly large effect on the strength of the operators as the contributions from the warped geometry far outweighed any running effects. The resultant wavefunction overlap of the six quark fields in the bulk play the most significant role in the suppression of the effective operators. The resultant overlap is sensitively controlled by the $c$ parameters of the quark fields which one determines by fitting the quark masses and the CKM parameters. 
The reason that these effective operators receive greater suppression than their proton decay counterparts stems from the simple fact that the $n$-$\overline{n}$ transition operators contain more fermion fields which leads to more negative contributions within the exponential overlap. This same simple reasoning should play a significant role in our intuition about effective operators of even higher mass dimension which are constructed from light fermion fields. The more light fermion fields that are present in the effective operator the more negative contributions we can expect within the resultant exponential wavefunction overlap. 
The significance of this result is that it shows that the geometric suppression from the warped RS background is sufficient to suppress the $n$-$\overline{n}$ transition operators to within the current experimental limits without any fine tuning while the effective mass scale can be as low as a fraction of a TeV. Any baryon number violating physics that may take place at a scale much higher than the electroweak symmetry breaking scale in the 5D theory should therefore get warped down to the TeV scale in the 4D effective theory.

\end{document}